\documentclass[
aps,%
12pt,%
final,%
notitlepage,%
oneside,%
onecolumn,%
nobibnotes,%
nofootinbib,%
superscriptaddress,%
noshowpacs,%
centertags]%
{revtex4}
\usepackage[russian]{babel}
\begin{document}
\selectlanguage{russian}
\title{New Estimates of Time Delays \\in the Gravitationally Lensed System PG1115+080} 
\author{\firstname{E.}~\surname{Shimanovskaya}}
\email{eshim@sai.msu.ru,lenashima@mail.ru}
\affiliation{%
Moscow MV Lomonosov State University, Sternberg Astronomical Institute, \\
Moscow 119992, Russia
}%
\author{\firstname{V.}~\surname{Oknyanskiy}}
\email{oknyan@sai.msu.ru}
\affiliation{%
Moscow MV Lomonosov State University, Sternberg Astronomical Institute, \\
Moscow 119992, Russia
}%
\author{\firstname{B.}~\surname{Artamonov}}
\email{artamon@sai.msu.ru}
\affiliation{%
Moscow MV Lomonosov State University, Sternberg Astronomical Institute, \\
Moscow 119992, Russia
}%
\begin{abstract}
We analyze all publicly available long-term optical observations of the gra\-vi\-ta\-tionally lensed quasar PG1115+080 for the purpose of estimating time delays between its four components. In particular, the light curves of PG1115+080 components obtained in 2001-2006 at Maidanak observatory (Uzbekistan) \cite{Tsvetkova2010} are considered. We find that the linear trend is observed in 2006 in light curves of all four components with fast variations only in the A1 and C components that can be due to microlensing and observational errors. Application of the MCCF method \cite{Oknyan1993} to the photometric data obtained in 2004-2005 gives values of time delays $\tau_{BC} = 22^{+2}_{-3}$, $\tau_{AC} = 12^{+2}_{-1}$ and $\tau_{BA} = 10^{+2}_{-3}$ days, which are in agreement with the results received earlier by Schechter \cite{Schechter1997} and Barkana \cite{Barkana1997} for 1995-1996 light curves with two different methods of statistic analysis. However, our estimates of $\tau_{BA}$ and $\tau_{BC}$ differ from the values received by the group of Vakulik based on the same Maidanak data \cite{Vakulik2009}. The ratio $\tau_{AC}/\tau_{BA}$ is equal to $\sim 1.2$ that is close to the value, received by Barkana ($\sim 1.13$) and predicted by lens models ($\sim1.4$), unlike the values received by Schechter ($\sim 0.7$) and Vakulik ($\sim 2.7$).
\end{abstract}
\maketitle


\section{Introduction}

According to General relativity, beams of light deviate from the straight line under the influence of the gravitational field of a massive object \cite{Einst1911}. If a galaxy is located close to the line of sight between a far quasar and an observer, the light from the quasar passes different ways that leads to formation of several images of the quasar. The internal variability of a quasar manifests itself in light curves of all these images, but with time shifts.  The time delay between manifestation of the internal variability of a quasar in a couple of its images is caused by geometrical difference of ways, light pass for each image, and difference of the gravitational potential in points corresponding to positions of quasar images.

Even before the discovery of the first gravitational lens, the double quasar QSO 0957+561, Refsdal showed that the time delay between manifestations of flux changes of a source (a supernova or a quasar) which is located behind a remote lensing galaxy and close to the line of sight allows determining the Hubble constant $H_0$ for certain model of the gravitational potential of a lensing galaxy \cite{Refsdal1964, Refsdal1966}. And vice versa, it is possible to investigate models of a lensing galaxy mass distribution using measured time delays and an estimated value of the Hubble constant, obtained with other methods.

A number of discovered gravitational lenses is increasing and every new one can become a candidate for Hubble constant estimation that allows minimizing systematic errors. By now, delays between components are measured for more than 20 lenses. Difficulties of time delay measurement in gravitational lens systems are concerned with the fact that the majority of gravitationally lensed quasars (GLQ) has rather small internal variability and there are manifestations of the micro-lensing effect due to separate stars of a lensing galaxy passing near ways of light forming images of a quasar. Therefore long-term homogeneous observations of GLQs with high resolution and robust time delay measurement methods are needed.

In this work, we analyze light curves of the components of the gravitationally lensed quasar PG 1115+080 to determine time delays. This is the second GLQ discovered and the first quadruple one. During first observations, three components  A, B and C with relatively large angle separation (about 2") were discovered \cite{Weymann1980}. Later the brightest component A was resolved into two components A1 and A2 spaced 0."48 apart \cite{Hege1981}, that  confirmed assumptions by Young et al \cite{Young1981}. The system configuration corresponds to the source position near a macrocaustics fold. The redshift of the source is $z_S = 1.722$ and the redshift of the lensing galaxy is $z_G = 0.31$ \cite{Henry1986, Christian1987, Tonry1998}. Thanks to its well-studied geometry \cite{Kristian1993, Courbin1997}, evident source variability and theoretically estimated time delays not exceeding dozens of days \cite{Vanderriest1986, Narasimha1992}, this system became the second GLS that was used to estimate the Hubble constant \cite{Schechter1997, Keeton1997}.

Long-term observation sets are needed to determine time delays. Such time series for PG1115+080 were first published by Schechter \cite{Schechter1997} in 1997. The quasar was observed in V band from November 1995 to June 1996. The Press method \cite{Press1992}, used to estimate time delays, showed that flux variations of $B$ component follow flux variations of $С$ component with the delay of $23.7\pm3.4$ days, and flux variations of $A1$ and $A2$ follow flux variations of $C$ with the delay of $9.4\pm3.4$ days. It is notable that time delay between A1 and A2 is less than a day. X-ray observations allowed estimating $\tau_{A_1 A_2}=0.149\pm 0.004$ days \cite{Chartas2003}. Therefore the light curves of $A_1$ and $A_2$ are usually combined for time delay estimation. 

The same data were analyzed by Barkana \cite{Barkana1997} with different statistical approach. Using analytical representation of the light curves and taking into account correlated errors of photometry, he obtained $\tau_{CB}=25.0^{+3.3}_{-3.8}$ days, which is in agreement with Schechter's estimate within the accuracy of the analysis, and $\tau_{CA}=13.4^{+2.0}_{-2.2}$ days, which is much larger then Schechter's value.
Another method of time delay estimation was applied to the same data set by Pelt et al. The time delays calculated with the minimum dispersion method are: $\tau_{AB} = 15.5 \pm 1.8$, $\tau_{CA}=10.3 \pm 1.9$, $\tau_{CB}=25.8 \pm 2.4$ days \cite{Pelt1998}.

New estimates of time delays in PG1115+080 were published in 2009 based on the analysis of images obtained with the 1.5-m telescope of the Maidanak Observatory (Uzbekistan) \cite{Vakulik2009} in 2004-2006. The authors applied their time delay estimation method and reported $\tau_{CB}=16.4^{+3.5}_{-2.5}$ and $\tau_{AC}=12.0^{+2.5}_{-2.}$ days. Those values substantially differ from Schechter's and Barkana's results and may suggest higher value of the Hubble constant.

We analysed the same Maidanak data sets with our modification of the Gaskell and Sparke method of the cross-correlation analysis \cite{Gaskell1986,GP1987,Oknyan1993} and obtained results supporting 'old' values of Schechter and Barkana \cite{Artamon2011}.  

Another attempt to measure time delays in PG1115+080 based on observations in 1995-1996 was undertaken in 2011 by Eulaers and Magain using two fundamentally different techniques \cite{Eulaers2011}. Modified numerical model fit technique \cite{Burud2001} resulted in $\tau_{CB} = 23.8^{+2.8}_{-3.0}$ and $\tau_{CA} = 11.7\pm2.2$, and minimum dispersion technique \cite{Pelt1996} resulted in $\tau_{CB} = 17.9 \pm 6.9$ and $\tau_{CA} = 7.6 \pm 3.9$ days. Because of such difference they concluded that estimation of time delays based on those data is impossible. 

Recently the Maidanak observations were re-analysed by Rathna Kumar еt al. using the difference-smoothing technique that resulted in $\tau_{CB}=13.2 \pm 9.0$ and $\tau_{CA} = 18.3 \pm 4.4$.

As the discrepancy of time delays obtained with different techniques has significant impact on the Hubble constant estimation, we aimed to more thoroughly inspect available observational data and apply our technique of time delay estimation.
In the following sections, we analyze available time series for PG1115+080, describe our modification of the Gaskell and Sparke method of the cross-correlation analysis \cite{Gaskell1986,GP1987,Oknyan1993} and discuss results of its application to publicly available observational data sets.

\section{Observations and reduction}

\begin{figure}
\setcaptionmargin{5mm} \onelinecaptionstrue
\includegraphics[width=3.5in]{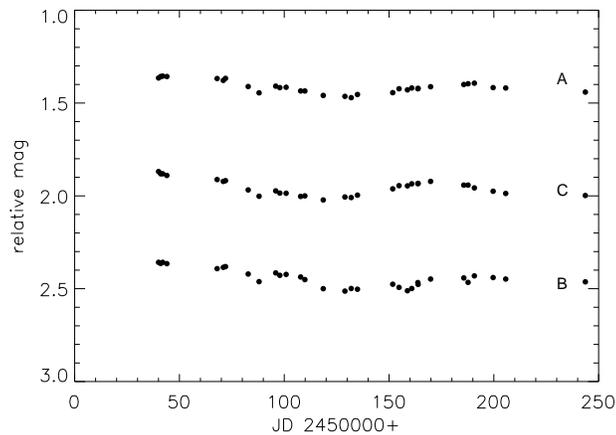}
\captionstyle{normal} \caption{The light curves of PG1115+080 components in 1995-1996 based on Schechter's data; the A light curve (averaged A1 and А2 light curves) is shifted by $1.5 mag$}\label{lc_s}
\end{figure}

There are three available long-term monitoring data sets for PG1115+080 suitable for time delay estimation:

S1) V band observations from November 1995 to June 1996 with four instruments: Hiltner 2.4 m telescope, the Wisconsin-Indiana-Yale-NOAO 3.5 m telescope (WIYN), the Nordic Optical 2.5 m telescope (NOT),
and the Du Pont 2.5 m telescope \cite{Schechter1997}. The light curves are presented in Fig.~\ref{lc_s}.

\begin{figure}
\setcaptionmargin{5mm} \onelinecaptionstrue
\includegraphics[width=3.5in]{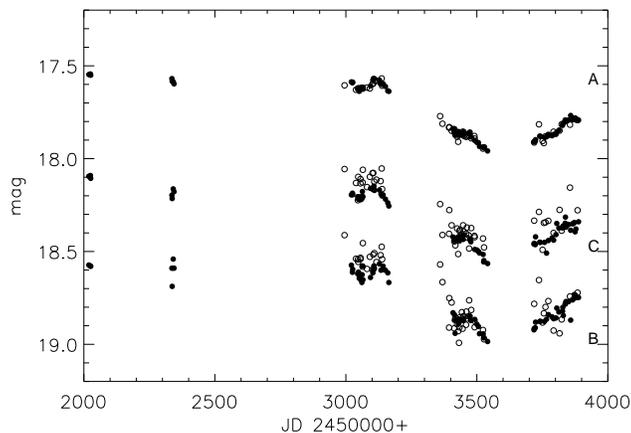} 
\captionstyle{normal} \caption{The light curves of PG1115+080 components in 2001-2006 (hollow circles -- CTIO+MDM data, filled circles -- Maidanak data)}\label{lc_m}
\end{figure}

S2) R band observations with the SMARTS 1.3 m telescope at CTIO and the 2.4 m telescope at the MDM Observatory in 2004-2006 \cite{Morgan2008}. The light curves of S2 are presented in Fig. \ref{lc_m} with hollow circles.

S3) R band observations with the 1.5 m telescope at Maidanak observatory in 2001-2006 \cite{Tsvetkova2010}.
The light curves of S3 are presented in Fig. \ref{lc_m} with filled circles.

The time delay between A1 and A2 spaced 0."48 apart is about several hours \cite{Chartas2003}, therefore we average light curves of A1 and A2 and compose the combined light curve A.

The PG1115+080 variability in 1995-1996 (S1) is less than $0.2\,mag$, moreover authors themselves call their photometry data preliminary and note the presence of correlated errors, probably, resulting from the PSF determination error. In 2004-2006, the quasar variability amplitude was a bit larger then in 1995-1996 observation season and reached $0.4\,mag$ \cite{Vakulik2009}. In the CTIO+MDM light curves (S2), the scatter in the data is too high, so the internal quasar variability becomes almost indistinguishable except A1 and A2 components. Mai\-da\-nak observations (S3) are more complete and homogeneous data of long-term monitoring of the object.


To combine data from different instruments and in different bands, we do as follows:

1) We convert all data from magnitudes to relative fluxes. It doesn't affect the position of the maximum of the cross-correlation function.

2) Then for each season we subtract average flux level from data to remove inter-season trends. It doesn't affect the position of the peak of the cross-correlation function either, but essentially affects its width making it smaller and hence increasing accuracy of time delay determination, as linear trends and slow inter-season changes lead to high positive correlation in a wide range of delays (hundreds of days and more) and prevent us from obtaining time delays with the required accuracy.
Additionally, this alignment procedure allows exclusion of the trend caused by slow microlensing.

\section{Statistical analysis of the PG 1115+080 light curves with the MCCF technique}

There are traditional cross-correlation techniques of the time delay estimation for evenly spaced time series. However astronomers usually have to deal with unevenly spaced time series. The problem of unevenness is solved in different methods in different ways. One can use linear or other interpolation and transform time series into evenly spaced ones and then use the classical procedure of the cross-correlation analysis. In doing so, the interpolation errors are introduced into initial data. Those errors can be mitigated if only one time set is interpolated leaving the other time set unchanged. A typical example of such an approach is the interpolation method of Gaskell and Sparke, hereafter called ICCF \cite{Gaskell1986, GP1987}. In the discrete correlation function method of Edelson and Krolik , hereafter called DCF \cite{Krolik1988}, the authors try to avoid interpolation, but they have to smooth and appro\-xi\-mate the cross-correlation function itself. Another class of methods involves construction of some continuous curve, approximating quasar variability light curve, and fitting it to both observational light curves through the minimization of the $\chi^2$ function to find parameters of that model curve and a time delay.

The variety of techniques used for estimation of time delays in the PG1115+080 and mentioned in the Introduction is due to the need to work at the limit of a method resolution. Our aim is to use available deficient information as accurately as possible.

\subsection{The MCCF technique}

To analyze available light curves of PG1115+080, we use method that is intermediate between ССF and DCF and is based on linear interpolation of the light curves \cite{GP1987}. 
Let $A(t_i)$ and $B(t_i)$ be the measured fluxes of two components of a GLS at a time moment $t_i$. The correlation coefficient between two time series $A(t)$ and $B(t)$ is calculated as follows:
\begin{equation}
      MCCF(\tau) =
      \frac{1}{M}\sum_{i,j}\frac{(A_{i}-\overline{A})(B_{j}-\overline{B})}{\sigma_{A}\sigma_{B}}\,,
      \label{CF}
   \end{equation}
where $M$ is the number of data pairs $(A_{i},B_{j})$ for every
$\tau-\Delta t_{max}\leq\Delta t_{ij}<\tau+\Delta t_{max}$ ($\Delta t_{ij}$ is a time lag between $t_{i}$ of the time series $A$ and $t_{j}$ of the time series $B$),
$\sigma_{x}$ is a standard deviation, $\overline{x}$ is the mean of $x$.
As the time series are unevenly spaced, we have to interpolate data at least in one of the light curves. If both time series have the same photometric accuracy and observations density, then initially we match the light curve $A(t)$ with the interpolated light curve $B(t+\tau)$ and calculate the cross-correlation coefficient for every $\tau$, next the light curves are interchanged, i.e. the observed light curve $B(t)$ is matched with the interpolated light curve $A(t+\tau)$. Thereby correlation coefficients are calculated twice for every time delay value, then average values of cross-correlation coefficients are calculated.

We modified the ICCF method by adding a restriction on usage of interpolated data: light curves are interpolated only at points which are sufficiently close to time of actual observations. The interval $\Delta t_{max}$ defining such proximity is given a priori and is selected according to the characteristic time of the quasar internal variability and the need for trade-off between interpolation errors and a sufficient number of pairs of data points to calculate the cross-correlation coefficient. Unlike the ICCF technique, where the number of pairs of points in calculation of the cross-correlation coefficient is equal for all time shifts, in MCCF this number varies, as we neglect some of the data points. All points are used in the analysis only for the zero delay and when a time delay does not exceed $\Delta t_{max}$. For time delays exceeding $\Delta t_{max}$, the number of pairs are always less than the maximum one. Yet we need a sufficient number of pairs for the significant estimate of the cross-correlation coefficient. If one choose too low $\Delta t_{max}$, the number of pairs of data points for calculation of the correlation coefficient for some time shifts may be insufficient. If the number of pairs for some time delays is less than ten ($M < 10$), we exclude the obtained value of the correlation coefficient from consideration. If one increase the interval $\Delta t_{max}$, the number of pairs for analysis grows, but our method tends to the ICCF method, i.e. the contribution of interpolation errors grows.

The ICCF technique implies not only interpolation, but also extrapolation: cross-correlated time series are extended at both edges with constant values equal to the first and last observation values. That means adding a large number of pairs of points not containing useful information, but smoothing the cross-correlation function, because inconclusive low-frequency part of the time series power spectrum is intensified. We extrapolate light curves beyond observation seasons only within the interval $\Delta t_{max}$. Those modifications was suggested by Oknyanskiy \cite{Oknyan1993}, and the modified method was called MCCF (Modified method of Cross-Correlation Function).

Light curves of GLQ components are irregular time series, but measurements of fluxes for all components are performed in the same days and in the same frames.
Therefore correlated errors can enhance correlation near zero delay or weaken it if the GLQ components are closely spaced and there is a systematic anti-correlation.
Some techniques, particularly the DCF one, propose to exclude data points obtained in the same night from the analysis. Unfortunately, we can not afford it, because then we have absolutely no data at the zero lag.
Stability and robustness of the MCCF technique is repeatedly confirmed: it has been successfully applied to determine the time delay between the components of the first gravitational lens QSO0957+561 \cite{Slavcheva2001}, 
and to determine the time delay between flux changes in the near infrared and optical regions of the spectrum for several active galactic nuclei \cite{Oknyanskij1999}, and for measuring time lags between flux changes in the optical continuum and emission lines, near-IR and optical continuum for NGC 4151 \cite{Oknyanskij1994, Oknyan1997, Oknyan2014}.
It is also used to determine the delay in the gravitational lenses
QSO2237+0305 (Einstein Cross) \cite{Koptelova2006} and UM673 \cite{Kop2012}.

\subsection{Application of the MCCF technique to light curves of PG1115+080}

Expected values of time delays in the gravitational lens PG1115+080 are of the order of dozens of days
 \cite{Narasimha1992}, that is significantly greater than the typical intervals between observations within a single season and less than an observation season duration, so chances of determining time delays, provided that there are an observable variability and lack of microlensing, are sufficiently large.

To successfully calculate a time delay, intervals between observations should not exceed the characteristic time of the variability of the object and the expected time delay. For the considered light curves of PG1115+080 it holds true. An observation season should be long enough so that there were actual variations of the brightness of the quasar. The time delay determination can be complicated due to the variability caused by microlensing by the stars of the galaxy-lens. This variability has a characteristic time of a few months, and can be observed in the form of slow inter-seasonal trends and changes. In any case, inter-seasonal changes will only hinder the estimation of the time delay.

In our case, when there are several observation seasons and gaps between seasons are significantly longer than the duration of the seasons, trends between seasons are not carry any useful information and lead to high correlation coefficients over the entire range of time delays and low relative amplitude of the main maximum of interest. 
Therefore we align the average flux of the seasons to the same level to avoid inconclusive correlation. It can be done because our interval of interest is small enough and points from neighbour seasons do not overlap when we shift light curves relative to each other by a time delay. 

\begin{figure}
\setcaptionmargin{5mm} \onelinecaptionstrue
\includegraphics[width=3.5in]{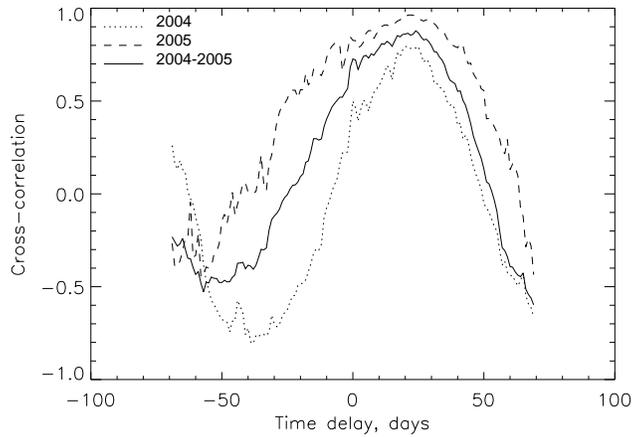}
\captionstyle{normal} \caption{Cross-correlation functions for the BC pair of components based on S3 (Maidanak) observations in 2004, 2005, and 2004-2005.}\label{mccf_bc20042005}
\end{figure}

\begin{figure}
\setcaptionmargin{5mm} \onelinecaptionstrue
\includegraphics[width=3.5in] {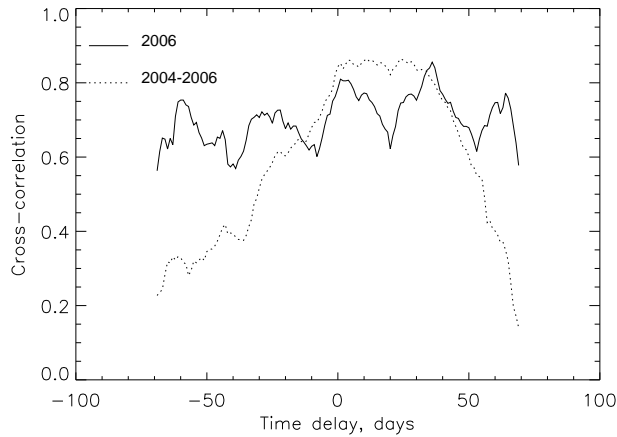}  
\captionstyle{normal} \caption{Cross-correlation functions for the BC pair of components based on the S3 (Maidanak) observations in 2006 and 2004-2006.}\label{mccf_bc20042006}
\end{figure}

The cross-correlation functions, calculated with the MCCF technique for the light curves of B and C components from Maidanak observations in 2004--2006, are presented in Figs. \ref{mccf_bc20042005} and \ref{mccf_bc20042006}.
The maxima of cross-correlation function for the seasons 2004, 2005 and 2004-2005 coincide with each other within the confidential interval, while the cross-correlation function for the 2006 has several peaks making it impossible to determine the maximum and therefore the time delay.

Impact of data obtained in 2006 leads to the fact that the cross-correlation function for 2004--2006 has no clearly defined maximum. So if one determines the position of the maximum of the cross-correlation function using the median value, this position will be shifted towards smaller delays. It is likely that the impact of 2006 data resulted in $\tau_{BC}=16.4$ days, calculated by  Vakulik team \cite{Vakulik2009}, that differs downwards from the values $\tau_{BC}=23.7$ and $\tau_{BC}=25.0$ calculated by Schechter and Barkana \cite{Schechter1997,Barkana1997}.

We noticed that the light curves of the A1, A2, B and C components in 2006
represent almost linear trend.
And in the A1 and C components, rapid oscillations are observed, which can be caused by observation errors.
We believe that these data can reduce the statistical significance of estimated time delay values and even lead to biased results. Therefore we decided to exclude the 2006 season from further statistical analysis.

\begin{figure}
\setcaptionmargin{5mm} \onelinecaptionstrue
\includegraphics[width=3.5in] {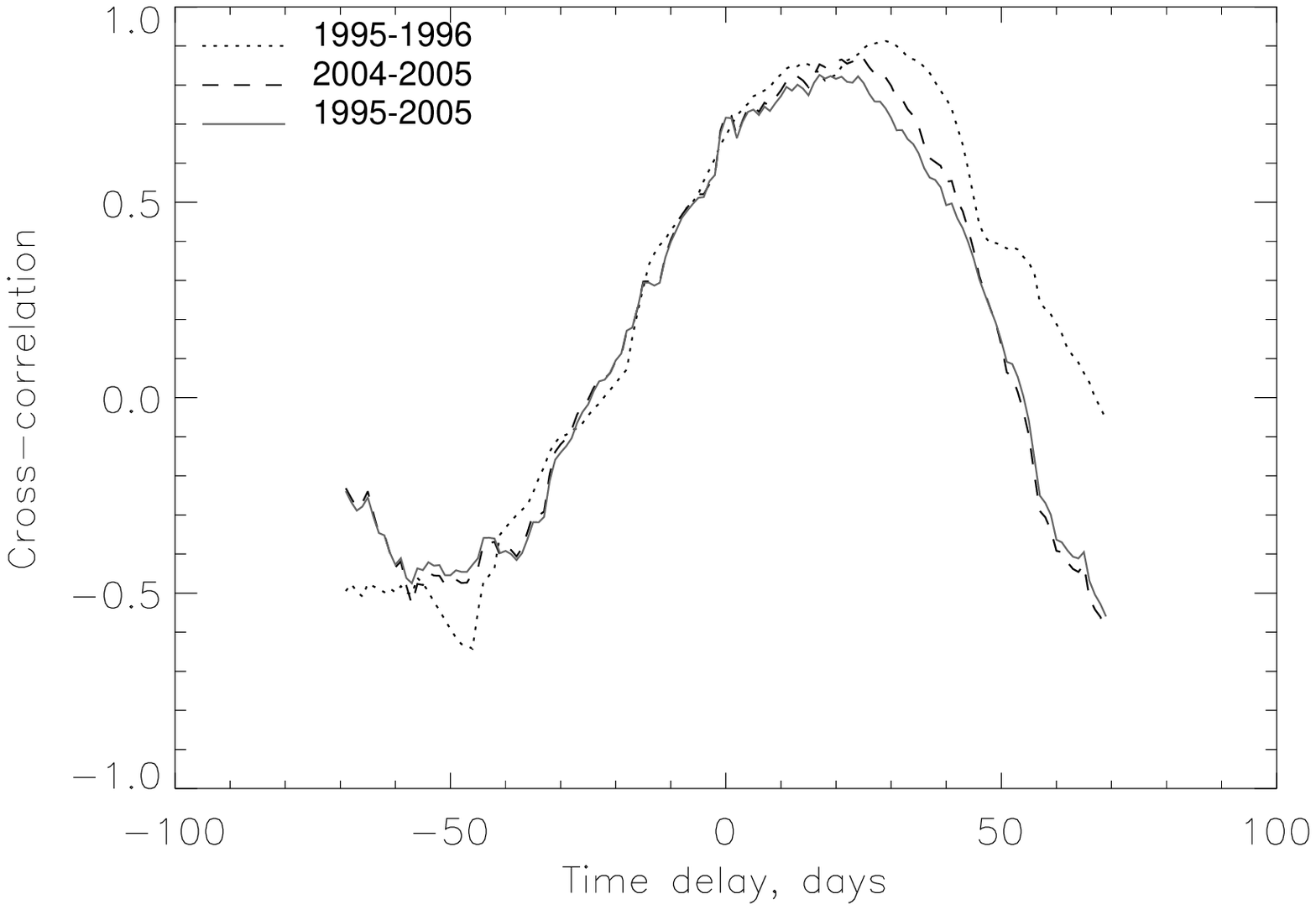} 
\captionstyle{normal} \caption{Cross-correlation functions for the BC pair of components based on the S1 data (dotted line), S3 data (dashed line), combined S1 and S3 data (continuous gray line)}\label{BC3grafst}
\end{figure}

Fig. \ref{BC3grafst} shows cross-correlation functions calculated for the B and C light curves in 1995-1996, 2004-2005, and the combined 1995-2005 light curves.
It can be seen here that the position of the maximum of the cross-correlation function, and hence the delay $\tau_{BC}$ based on observations in 1995-1996, is greater than based on observations in 2004-2005.
The combined 1995-2005 light curves also yield the time delay value $\tau_{BC}= 22$ days. This is the same value as based on observations in 2004-2005, as this observation season provides more pairs of points for the calculation of a cross-correlation coefficient.

\begin{figure}
\setcaptionmargin{5mm} \onelinecaptionstrue
\includegraphics[width=3.5in] {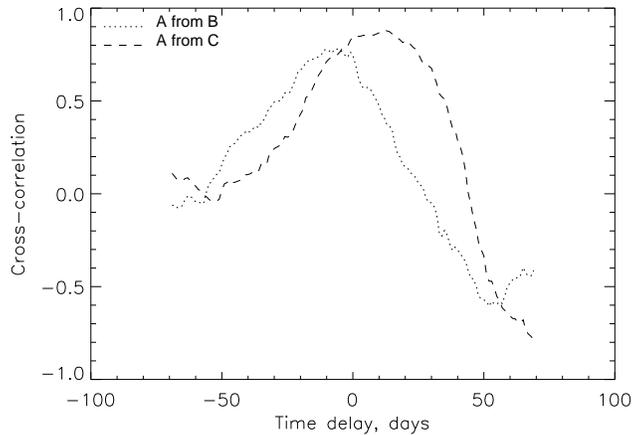}
\captionstyle{normal} \caption{Cross-correlation functions for the AB and AC pairs of the
components  based on S3 (Maidanak) observations in 2004-2005.}\label{mccf_abac}
\end{figure}

The cross-correlation functions calculated for pairs AB and AC in 2004-2005 are presented in Fig. \ref{mccf_abac}.  
Positions of the peaks correspond to the delay of the internal quasar variability manifestation in A component relative to B component $\tau_{BA}= 10$ days and relative to C component $\tau_{AС}= 12$ days.

\begin{figure}
\setcaptionmargin{5mm} \onelinecaptionstrue
\includegraphics[width=3.5in] {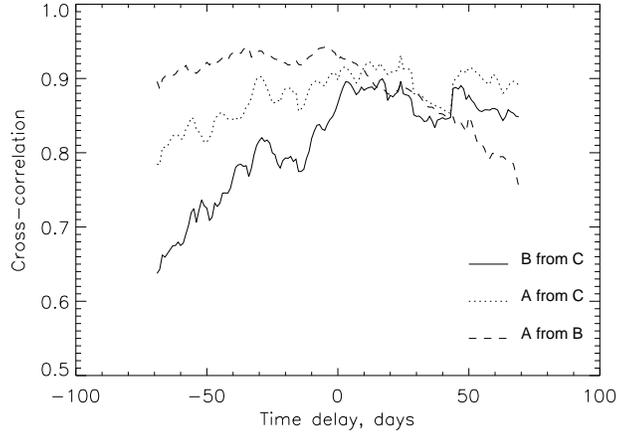}
\captionstyle{normal} \caption{
Cross-correlation functions for the AB, AC, and BC pairs of the
components based on S2 (CTIO+MDM) observations in 2004-2006.}\label{mccf_cbabac}
\end{figure}

The CTIO + MDM data (S2) overlap with Maidanak data (S3), but the quasar internal variability is practically indistinguishable in light curves of B and C components due to the large scatter of the data points.
Cross-correlation functions for the data shown in Fig.~\ref{mccf_cbabac} do not allow determining the time delays.
Vakulik and coauthors also reported that is was rather problematic to use the CTIO + MDM data to determine the time delays with their method.
We have tried to include the light curve of the A component from S2 into the analysis in combination with Maidanak data, but it does not affect obtained time delay values.

\subsection{Uncertainties}

To calculate the uncertainty of time delay determination, we use Monte-Carlo approach that means application of the MCCF technique to a statistically large number of simulated light curve pairs with the specified time delay.
The simulated light curves have to reflect the statistical properties of the quasar variability and properties of the observed light curves.
For example, one of the ways to assess robustness of the method is to simulate the light curves from the real data and compare the light curves with some points removed by chance \cite{Pelt1994}. However that approach underestimates time-delay uncertainties because not all statistical properties of quasar variability are incorporated into the simulated light curves. Therefore we decided to construct every pair of light curves based on a simulated light curve of the quasar variability with randomized parameters. The second light curve is produced by shifting the first one by the specified time delay. Then we select points according to actual observations moments.

The quasar PG1115+080 can be referred to the family of quasars with average statistical properties.
A quasar power spectrum in optical band can be modelled as power law with a mean index
$\alpha = -1$ \cite{Zheng1997}.
To simulate the quasar light curve, we need an algorithm that is able to produce time series with the power-density spectrum $S(\omega)\sim \omega^\alpha$ randomizing both the phase and the amplitude of the Fourier transform of a signal. Such an algorithm was proposed by Timmer\& Konig \cite{Timmer1995}.

The first step is to generate the power-density spectrum:
\begin{equation}
S(\omega)=N(0,1)\sqrt{\frac{1}{\omega}}+i N(0,1)\sqrt{\frac{1}{\omega}}
\end{equation}
where $N(0,1)$ is a normally distributed stochastic variable with
zero mean, and the standard deviation equal to $1$.
The simulated light curve is produced by applying the inverse Fourier transform to that power spectrum followed by renormalizing to the observed magnitudes. The second light curve is generated by shifting the simulated light curve by given time delay. To reflect the unevenness of actual light curves, the generated light curves are resampled as the observational ones. To simulate the photometry uncertainty, normally distributed random variables with zero mean and standard deviation equal to the actual photometry errors are added to the simulated light curves.

\begin{figure}
\setcaptionmargin{5mm} \onelinecaptionstrue
\includegraphics[scale=1]{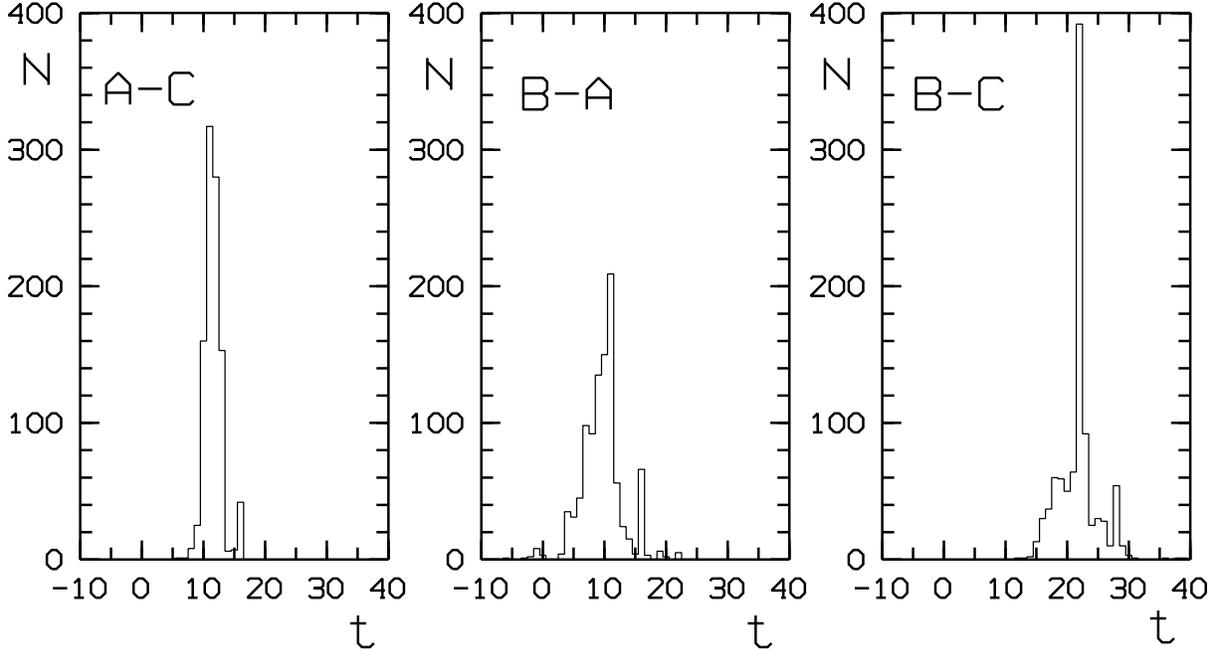}  
\captionstyle{normal} \caption{
Histograms of time delays recovered from cross-correlation
analysis of 1000 Monte Carlo simulations of the A and C, B and A, B and C light curves.}\label{montecarlo}
\end{figure}

The Monte-Carlo histograms are presented in fig. \ref{montecarlo}. We assume the range where 68 \% of the 1000 realizations falls to be a confidence interval of the time delay determination ($1\sigma$).

\section{Discussion}

Our results of time delay determination for PG1115+080 are summarized in the last row of the Table~\ref{tdtab}. The uncertainties correspond to $1\sigma$. Time delay values, calculated by other researchers, are also presented in the table for comparison.

\begin{table}
\setcaptionmargin{0mm} \onelinecaptionstrue
\captionstyle{flushleft} \caption{Time delays between images of the lensed quasar PG1115+080 estimated with different techniques.}
\bigskip
 \begin{tabular}{l|c|c|c|c|c}  \hline \hline
References                                 & Season   & $\tau_{BA}$  & $\tau_{AC}$    & $\tau_{BC}$  & $\tau_{AC}/\tau_{BA}$ \\
                                       &                  & (days)    & (days)      & (days)    &   \\
\hline
P. Schechter et al \cite{Schechter1997} &  & $14.3\pm 3.4$ & $9.4\pm 3.4$ & $23.7\pm 3.4$ & $0.7 \pm 0.4$ \\
R. Barkana \cite{Barkana1997}          & 1995--1996 & $11.7^{+2.9}_{-3.2}$ & $13.3^{+2.0}_{-2.2}$ & $25.0^{+3.3}_{-3.8}$ &  $1.13^{+0.18}_{-0.17}$\\
J. Pelt et al \cite{Pelt1998}  & & $15.5 \pm 1.8$ & $10.3 \pm 1.9$ & $25.8 \pm 2.4$ & $0.7 \pm 0.2$\\
E. Eulaers \& P. Magain \cite{Eulaers2011} NMF & & & $11.7\pm 2.2$ & $23.8^{+2.8}_{-3.0}$ &  \\
E. Eulaers \& P. Magain \cite{Eulaers2011} MD & & &  $7.6 \pm 3.9$ & $17.9  \pm 6.9 $     &  \\
\hline
V. Vakulik et al \cite{Vakulik2009}    & 2004--2006 & $4.4^{+3.2}_{-2.5}$ & $12.0^{+2.5}_{-2.0}$ & $16.4^{+3.5}_{-2.5}$ & $2.73^{+2.55}_{-2.01}$ \\
S. Rathna Kumar et al. \cite{Kumar2014} &  & $8.9 \pm 8.4$ &  $18.3 \pm 4.4$ & $13.2 \pm 9.0$     &  $2.1 \pm 2.5$\\
\hline
E. Shimanovskaya et al. \cite{Shima2015} & 2004--2005 & $10^{+2}_{-3}$ & $12^{+2}_{-1}$ & $22^{+2}_{-3}$ & $1.2^{+0.4}_{-0.5}$    \\
\hline
\label{tdtab}
\end{tabular}
\end{table}

Our estimate of $\tau_{BA}$ match the results of Schechter and Barkana within the accuracy of the analysis, but differs greatly from $\tau_{BA}$ calculated by Vakulik et al. for the same Maidanak data except 2006 season that we excluded from the analysis. As for $\tau_{AC}$, almost all techniques yield similar time delay approximately 12 days. Only recent Rathna Kumar' result is a bit larger. The time delay $\tau_{BC}$, obtained by Vakulik et al. and Rathna Kumar et al., is less than $\tau_{BC}$, calculated by Schechter and Barkana and confirmed with our MCCF technique, by a factor of 1.5.

In the last column of the Table \ref{tdtab}, the ratio $\tau_{AC}/\tau_{BA}$ for PG1115+080 is presented.
The time delay ratio doesn't depend on the Hubble constant and can be used for lens models comparison. Keeton and Kochanek investigated different models of the PG1115+080 lensing galaxy and theoretically calculated the ratio $\tau_{AC}/\tau_{BA}$ to be in the range from $1.35$ to $1.47$ \cite{Keeton1997}. As can be seen in the last column of the Table \ref{tdtab}, the Schechter's value $\tau_{AC}/\tau_{BA}$ is significantly less than model predicted values, the ratio calculated by Barkana is close to the model predicted range, Vakulik's value is more than twice as large as the theoretically predicted value and has the large uncertainty. Our value $\tau_{AC}/\tau_{BA}=1.2^{+0.44}_{-0.46}$, is close to the model predicted value.

Based on measured time delays in five lenses (RXJ0911+0551, PG1115+080, SBS1520+530, B1600+434, and HE2149-2745), Kochanek estimated $H_0 = 48^{+7}_{-4} \, km\, s^{-1}\, Mpc^{-1}$, if a lensing galaxy has isothermal mass distribution, and $H_0 = 71\pm 6 \, km\, s^{-1}\, Mpc^{-1}$, if a lensing galaxy has constant mass-to-light ratio \cite{Kochanek2002}. Recently, the values $H_0= 69 \pm 6^{stat} \pm 4^{syst} \, km\, s^{-1}\, Mpc^{-1}$ \cite{Sereno2014} and  $H_0= 71.8 \pm  5.8 \, km\, s^{-1}\, Mpc^{-1}$ \cite{Kumar2014} were derived based on lightcurves of 18 gravitationally lensed quasars. Those $H_0$ values agree with a recent estimate of the Hubble constant based on Cepheids observations in the HST Key Project $H_0=73.8 \pm 2.4\, km\, s^{-1}\, Mpc^{-1}$ \cite{Riess2011}. Combining WMAP and SDSS CMB spectrum observations with the assumption of the flat Universe gives value $H_0=70^{+0.04}_{-0.03}\, km\, s^{-1}\, Mpc^{-1}$ \cite{Tegmark2004}. New results of the ESA's Planck project testify to the smaller speed of expansion of the Universe $H_0=67.3\pm 1.2\, km\, s^{-1}\, Mpc^{-1}$ \cite{Plank2013}, however in some recent works this result is called into question because of frequency channel calibration issues of the Planck telescope \cite{Spergel2013}.  Another method of the Hubble constant estimation based on observations of Syunyaev-Zeldovich effect gives $H_0= 59^{+10^{rand},+8^{syst}}_{-9^{rand},-7^{syst}}\, km\, s^{-1}\, Mpc^{-1}$ for standard cold dark matter with $\Omega_M= 1.0$, $\Omega_\Lambda= 0.0$ or $H_0= 66^{+11,+9}_{-10,-8}\, km\, s^{-1}\, Mpc^{-1}$ if $\Omega_M= 0.3$, $\Omega_\Lambda= 0.7$ \cite{Jones2005}.
Investigation of lens models shows that the $H_0$ estimated based on gravitational lensing is close to values obtained with other methods only if a lensing galaxy contains small or zero portion of dark matter \cite{Impey1998}. Hence, time delays between gravitational lens components allow estimating not only Hubble constant value, but portions of dark energy and dark matter also.

\begin{table}
\setcaptionmargin{0mm} \onelinecaptionstrue
\captionstyle{flushleft} \caption{Recent Hubble constant estimates.}
\bigskip
 \begin{tabular}{l|c|c}  \hline \hline
Method    &  $H_0$, $km\, s^{-1}\, Mpc^{-1}$   &  References  \\ 
\hline
Cepheids (HST) & $73.8 \pm 2.4$ ($3.3 \%$) &  Riess et al 2011 \cite{Riess2011} \\
CMB: WMAP+SDSS,  &  $70 \pm 4$ ($5.7 \%$) & Tegmark et al 2004 \cite{Tegmark2004} \\
WMAP7, & $71.0 \pm 2.5$ ($3.5 \%$)  & Jarosik et al 2011 \cite{Jarosik2011} \\
PLANCK & $67.3\pm 1.2$ &  Planck Collaboration 2014 \cite{Plank2013} \\
Sunyaev-Zeldovich effect & $66 \pm 11$ ($16.7\%$) & Jones et al 2005 \cite{Jones2005} \\
AGN IR lags & $73 \pm 3$ & Yoshii et al. 2014 \cite{Yoshii2014} \\
Gravitational lenses & $61 \pm 7$ ($11.5\%$)  & Courbin 2003 \cite{Courbin2003}\\
 & $68 \pm 6$ ($8.8 \%$) &  Oguri 2007 \cite{Oguri2007}\\
  & $69 \pm 6^{stat} \pm 4^{syst}$ ($15 \%$) &  Sereno \& Paraficz 2014 \cite{Sereno2014}\\
    & $71.8 \pm  5.8$ ($8 \%$) &  Rathna Kumar et al. 2014 \cite{Kumar2014}\\
\hline
\label{h0tab}
\end{tabular}
\end{table}

The time delay $\tau_{BC}$ is less affected by systematic errors in comparison with shorter delays $\tau_{BA}$ and $\tau_{AC}$. For this reason it is more often used for the Hubble constant estimation (e.g. see \cite{Schechter1997, Keeton1997}). Disregarding the whole variety of lens gravitation potential models, let's take into account just the fact that the time delay is inversely proportional to the Hubble constant \cite{Refsdal1964, Cooke1975, Kayser1983, Borgeest1983, Schneider1985}. The $H_0$ value, estimated based on $\tau_{BC}$ calculated by Vakulik et al. or Rathna Kumar et al., is 1.5-times greater \cite{Tsvetkova2010} and hence closer to values measured with other methods, e.g. observations of Cepheids. However our value of $\tau_{BC}$ doesn't support that result. Vise versa: it takes us back to the traditional situation when the Hubble constant estimates from time delays in gravitational lenses are significantly less than $H_0$ values measured with other methods. Notice that the value of the Hubble constant measured in the Planck program is less than the $H_0$ value derived from Cepheids observations as well \cite{Plank2013}.


\section{Conclusion}

At first glance, the problem of measuring time delays between manifestations of internal quasar variability in components of a GLQ seems quite simple. However, actual observations are far from ideal case of evenly spaced time series. Except for observations uncertainties and large gaps in the light curves requiring usage of interpolation, su\-per\-po\-si\-ti\-on of mic\-ro\-lensing and internal variability of the quasar and quite small amplitude of the quasar internal variability pose many problems. Even for such 'convenient' lens for time delay estimation as PG1115+080, different researchers obtain different values of time delays, as indicated in the Table~\ref{tdtab}. This is particularly due to work at the limit of resolution of time delay methods, lack of observational data needed for robust estimation of time delays, and some parts of data not containing useful information for time delay estimation, as in the case of 2006 season.

Different values of time delays lead to dissimilarity in the Hubble parameter estimation. Values of this universal constant calculated from time delays between GLS components are usually smaller than values obtained with other techniques. New estimates of time delays in the PG1115+080 gravitational lens, published by Vakulik et al, suggest larger value of the Hubble constant then previous estimates \cite{Tsvetkova2010} narrowing a gap between values obtained with different approaches.

We analysed all available optical observations of the gravitational lens PG1115+080 with our cross-correlation technique to confirm or argue against that result. We failed to determine time delays based on combined data because of inhomogeneity of observations and possible errors of their initial processing. Observations  of PG1115+080 in R band on 1.3-m SMARTS telescope (CTIO) and on 2.4-m telescope of MDM Observatory in 2004-2006 \cite{Morgan2008} are unsuitable for the time delay estimation with both Vakulik's \cite{Vakulik2009} and ours techniques, because the quasar internal variability is bleared by chaotic abrupt behavior of light curves that may be caused by microlensing and photometry errors. Among three available long-term data sets of optical observations of PG1115+080, the most complete and homogeneous ones were obtained at Maidanak observatory (Uzbekistan) in 2004-2006 \cite{Tsvetkova2010}. But we find that light curves of all four components of the GLQ in 2006 represents almost linear trends. Fast variations are observed only in A1 and C components that can be as due to pho\-to\-metry errors or microlensing. These fast variations can decrease statistical significance of the time delay estimates or even produce misleading results. Therefore we decided to exclude observations in 2006 from the cross-correlation analysis. Time delays estimated with the MCCF technique \cite{Oknyan1993} applied to Maidanak observations in 2004-2005 are in better agreement with previous results by Schechter \cite{Schechter1997} and Barkana \cite{Barkana1997} obtained for 1995-1996 light curves with two different techniques of statistical analysis than with new estimates by Vakulik et al \cite{Vakulik2009, Tsvetkova2010} and Rathna Kumar et al \cite{Kumar2013}. The ratio $\tau_{AС}/\tau_{BA}$ is in agreement with Barkana's value and model predictions supporting adequacy of our results. Our estimates of time delay suggest smaller value of the Hubble parameter than Vakulik et al results.

As long as different methods yield different values of time delays for the same data, it's early to talk about estimation of the Hubble constant. We have to identify reasons of such difference that can be due to interpolation and accounting for microlensing. To attain better accuracy, new homogeneous long-term observations and new methods of analysis taking into account microlensing and weight of every data point are also needed.

\textbf{Acknowledgements}

We would like to thank Paul L. Schechter for kindly providing us with the PG1115+080 observations data obtained in 1995-1996. This research is supported by Russian Foundation for Basic Research, grant 14-02-01274. We made use of NASA's Astrophysics Data System.
\appendix

\end{document}